\documentclass[twocolumn,amsmath,amssymb,superscriptaddress]{revtex4}

\usepackage{graphicx} % Include figure files
\usepackage{dcolumn}  % Align table columns on decimal point
\usepackage{bm}       % bold math

\begin{document}

%\preprint{APS/123-QED}

\title{Near degeneracy and pseudo Jahn-Teller effects in mixed-valence
ladders:\\ the phase transition in NaV$_{2}$O$_{5}$ }

\author{L. Hozoi}
\affiliation{Max-Planck-Institut f\"{u}r Festk\"{o}rperforschung,
             Heisenbergstrasse 1, 70569 Stuttgart, Germany }

\author{S. Nishimoto}
\affiliation{Institut f\"{u}r Theoretische Physik,
             G\"{o}ttingen Universit\"{a}t, Friedrich-Hund-Platz 1,
             37077 G\"{o}ttingen, Germany }

\author{A. Yamasaki}
\affiliation{Max-Planck-Institut f\"{u}r Festk\"{o}rperforschung,
             Heisenbergstrasse 1, 70569 Stuttgart, Germany }

\date{\today}

\begin{abstract}
We analyze the electronic structure of a mixed-valence ladder
system. We find that structural anisotropy and complex electron
correlations lead to on-rung charge localization and insulating
character. Charge fluctuations within the rung of the ladder
interact strongly to the lattice degrees of freedom, which gives
rise to large pseudo Jahn--Teller effects. The phase transition in
NaV$_2$O$_5$ should be driven by this kind of mechanism.
\end{abstract}

%\pacs{71.27.+a, 71.28.+d, 71.38.-k, 77.80.-e}
%\keywords{suggested keywords ?}

\maketitle

Stoichiometric compounds in which the formal valence state of an
element is fractional are referred to as mixed-valence (MV) systems.
MV materials display remarkable properties, such as
ferrimagnetism and a metal\,--\,semiconductor transition in
magnetite, a transition from a ferromagnetic metal
to an antiferromagnetic insulating state in some half-doped
manganites or the insulator\,--\,insulator transition towards a 
spin-gapped state in NaV$_2$O$_5$\,.
The phase transitions in these compounds are often discussed in
terms of charge ordering (CO) at the transition metal (TM) sites. 
In such models, electrostatic interactions should yield at low temperature (LT)
an ordered arrangement of the extra $d$ electrons, those
associated in the ionic picture with fractional oxidation states
for the TM cations. For NaV$_2$O$_5$ it was
initially believed that charge order is present at all
temperatures, in the form of alternating chains of V$^{4+}$ and
V$^{5+}$ \cite{carpy_navo}.
However, it was later found that above 34\,K the vanadiums are all
equivalent
\cite{smolinski_navo_98,navo_xrd_Pmmn} and some
kind of CO should only occur in the LT
phase, presumably with a zigzag arrangement of the $d$ electrons
\cite{navo_afe_99,seo_navo_CO_98,mostovoy_navo_00_02}.
In this Letter we investigate electron\,--\,lattice interactions in NaV$_2$O$_5$\,. We take explicitly into account the ligand degrees of freedom and find that most probably the phase transition at low temperature is induced by collective, pseudo Jahn--Teller (JT) effects involving TM--$3d$ and oxygen--$2p$ hole states localized within V--\,O\,--V clusters. 
Our findings indicate different physics as
compared to the commonly accepted $d$\,--\,$d$ CO model. 
Similar JT type effects, we believe, may be relevant 
for other MV or doped TM materials.

Corner-sharing [VO$_5$] pyramids form two-leg ladders in
NaV$_2$O$_5$\,. Such ladders are assembled in quasi
two-dimensional layers, with pyramids on adjacent ladders sharing
edges of their bases. 
Interatomic interactions are strongly
anisotropic, as shown by first principles band-structure
calculations \cite{smolinski_navo_98,yaresko_navo_00,navo_dmft_02}.
This is related to obvious structural features: V--O--V bond angles of
approximately 90$^{\mathrm{o}}$ between adjacent ladders, shorter V--O
bonds across the rung than along the leg, and tilting of the
[VO$_5$] pyramids towards the rung bridging oxygen
\cite{navo_xrd_Pmmn}. The insulating character is associated with this rung\,--\,leg anisotropy and relatively large inter-rung Coulomb interactions
\cite{smolinski_navo_98,seo_navo_CO_98,mostovoy_navo_00_02,yaresko_navo_00,navo_dmft_02,horsch_navo_98,satoshi_navo_gs_98}.
In a first approximation a single $d$ electron resides on each
rung, in an essentially non-bonding $d_{xy}$--\,$d_{xy}$ orbital,
where the $x$ axis is parallel to the rung and $y$ is along the
leg \cite{smolinski_navo_98,yaresko_navo_00}. Alternatively, in a configuration-interaction
like picture, a superposition of left and right occupied orbital
states, $d_{L,xy}^{\,1}d_{R,xy}^{\,0}$ and
$d_{L,xy}^{\,0}d_{R,xy}^{\,1}$\,, would correspond to each V--O--V
rung. If the energy difference between the V\,$3d$ and O\,$2p$
levels is not too large, O\,$2p^{5}$ configurations may contribute
as well to the ground-state (GS) wave-function. The ligand-hole
configurations with the largest weight would imply holes ``localized''
on the rung oxygen O$_{\mathrm{R}}$\,,
$d_{xy\,}^{\,1}p_{y\,}^{\,1}d_{xy\,}^{\,1}$\,, due to strongly
anisotropic $d_{xy}$--\,$p_{x,y}$ overlap integrals and lower
Madelung potential at the O$_{\mathrm{R}}$ site as compared to the leg
oxygens O$_{\mathrm{L}}$ \cite{2nd_note}. Density functional
calculations indicate an energy difference
$\Delta\!=\!\epsilon_{d}-\epsilon_{p}\!\approx\!4$\,eV between the
V\,$d_{xy}$ and O$_{\mathrm{R}}$\,$p_{y}$ levels. However, neglecting
inter-site Coulomb interactions, the on-site electron repulsion
should reduce this number to $\Delta_{\mathrm{eff}}\!=\!\Delta-U_{pp}$
for $p^{6}d^{0}$$\rightarrow$\,$p^{5}d^{1}$ charge transfer. Due
to the extended nature of the oxygen $2p$ functions there is no
unambiguous procedure of estimating the Coulomb repulsion $U_{pp}$
between two electrons in such an orbital. Nevertheless,
semiquantitative estimates in copper oxides suggest a value as
high as 4\,eV \cite{Upp_mcmahan_88}.

%% Table 1
\begin{table}[b]
\caption{In-plane NMTO orbital energies and hoppings for the
high-temperature structure of NaV$_2$O$_5$ (eV). O$_{\mathrm{L}}'$
stands for an oxygen ion on the leg of an adjacent ladder. For the
DMRG calculations we used 
$\epsilon_{y}^l\!=\!\epsilon_{x}^l\!=\!-4.9$\,.}
\begin{ruledtabular}
\begin{tabular}{rl}
$pd$ and $pp$ hoppings                        &Orbital energies \\
\colrule
 V\,$d_{xy}$--\,O$_{\mathrm{R}} $\,$p_{y}  $: $t_{r} $\,=\,1.3
&V: $\epsilon_{xy} $\,=\,0                                      \\
 V\,$d_{xy}$--\,O$_{\mathrm{L}} $\,$p_{x}$\,: $t_{l} $\,=\,0.8
&O$_{\mathrm{R}}$: $\epsilon_{y}^r$\,=\,--3.5                   \\
 V\,$d_{xy}$--\,O$_{\mathrm{L}}'$\,$p_{y}$\,: $t_{l}'$\,=\,0.7
&O$_{\mathrm{L}}$: $\epsilon_{x}^l$\,=\,--4.9                   \\
 O$_{\mathrm{R}}$\,$p_{y}$--\,O$_{\mathrm{L}} $\,$p_{x}$: $t_{rl}$=\,0.4
&O$_{\mathrm{L}}$: $\epsilon_{y}^l$\,=\,--5.2                   \\
 O$_{\mathrm{L}}$\,$p_{x}$--\,O$_{\mathrm{L}}'$\,$p_{y}$: $t_{ll}$\,=\,0.3\\
\colrule
 V\,$d_{xy}$--\,O$_{\mathrm{L}} $\,$p_{y}$\,: $t$\,=\,0.3             \\
 O$_{\mathrm{R}}$\,$p_{y}$--\,O$_{\mathrm{L}} $\,$p_{y}$: $t$\,=\,0.2
&O$_{\mathrm{L}}$\,$p_{y}$--\,O$_{\mathrm{L}}'$\,$p_{y}$: $t$\,=\,0.5 \\
 O$_{\mathrm{L}}$\,$p_{x}$--\,O$_{\mathrm{R}} $\,$p_{x}$: $t$\,=\,0.2
&O$_{\mathrm{L}}$\,$p_{x}$--\,O$_{\mathrm{L}}'$\,$p_{x}$: $t$\,=\,0.5 \\
\end{tabular}
\end{ruledtabular}
\end{table}

In Table\,I we reproduce hopping matrix elements and orbital
energies describing the $d_{xy}$--\,$p_{x,y}$ and
$p_{x,y}$--\,$p_{x,y}$ bonding on the V--O ladder.
These were obtained by applying a powerful technique, the
$N$th-order muffin-tin orbital (NMTO) downfolding method recently
developed in Stuttgart \cite{nmto}. In order to generate an
effective Hamiltonian capable of modeling metal\,--\,ligand
interactions, we downfold all basis functions other than
V\,$d_{xy}$ and O\,$2p$. 

We use the NMTO hoppings and orbital energies to
investigate the nature of the many-electron GS
wave-function of a V--O Hubbard type ladder model. The Hamiltonian
is written as \cite{satoshi_navo_gs_98,dmrg_CuO_98_02}:
\begin{eqnarray*}
H &= &\epsilon_d \sum_{i\sigma} n_{i\sigma}
     +\sum_{j\sigma} \epsilon_p^j\,n_{j\sigma} 
     +\sum_{<ij>\sigma} 
       t_{pd}^{j}(d_{i\sigma}^{\dagger}p_{j\sigma} + h.c.)      \\
  &+ &\sum_{<jj'>\sigma} 
       t_{pp}^{jj'}\,(p_{j\sigma}^{\dagger}p_{j'\sigma} + h.c.) 
     +U_{dd} \sum_i n_{i\uparrow}n_{i\downarrow}                \\
  &+ &U_{pp} \sum_j n_{j\uparrow}n_{j\downarrow} +
      V_{pd} \sum_{<ij>\sigma\sigma'} n_{i\sigma}n_{j\sigma'}  \ .
\end{eqnarray*}
We neglect the apical ligands and restrict to a single orbital at each site: V\,$d_{xy}$, O$_{\mathrm{R}}$\,$p_{y}$, O$_{\mathrm{L}}$\,$p_{x}$ on the same ladder plus nearest neighbor (n.n.)
O$_{\mathrm{L}}'$\,$p_{y}$ orbitals on adjacent ladders. 
We also neglect the finite V\,$d_{xy}$--\,O$_{\mathrm{L}}$\,$p_{y}$
overlap, which should be zero for V and O$_{\mathrm{L}}$ atoms in the
same $zy$ plane, and $\pi$-type $pp$ hoppings like those listed
in the lower part of Table\,I\,. 
Still, we believe that our model is able to capture the essential properties of V--O ladders in NaV$_2$O$_5$\,. 
GS expectation values are obtained by density-matrix renormalization-group (DMRG) methods \cite{dmrg_review_scholl}. For these calculations up to 20 rungs on the V--O ladder, open-end boundary conditions, and up to $m\!=\!2000$ states to build the DMRG basis were used. 
The typical discarded weight was $10^{-11}\!-\!10^{-10}$.
We employed the code developed by Jeckelmann \textit{et al.}\,\cite{dmrg_CuO_98_02}. 
The results reported here were obtained with $U_{dd}\!=\!3$\,eV, 
as deduced in 
\cite{smolinski_navo_98,yaresko_navo_00,navo_dmft_02}, and the 
same $U_{pp}$ for O$_{\mathrm{R}}$ and O$_{\mathrm{L}}$.

For reasonable values of the $U_{pp}$ and $V_{pd}$ parameters, the DMRG results show significant O$_{\mathrm{R}}$ hole character for the GS wave-function, see Fig.1(a). With $U_{pp}\!=\!4$\,eV and
$V_{pd}\!=\!0.25$\,eV, for example, the occupation of the rung $p_y$ orbital is 1.60, which indicates that the overall weights of the O$_{\mathrm{R}}$ $p_{y\,}^{\,2}$ and $p_{y\,}^{\,1}$ configurations are nearly equal. Yet the ladder is insulating, with a finite charge transfer gap, see Fig.1(b).
The fact that the gap is strongly dependent on the rung\,--\,leg anisotropy was previously evidenced in \cite{satoshi_navo_gs_98}.
Regarding the other parameters, the most important is $U_{pp}$ in our model. In simple words, for a given $V_{pd}$, the size of $U_{dd}$ ``matters'' when $U_{pp}$ is sufficiently high to give large occupation at the $d$ sites.

%% FIGURE 1
\begin{figure}
\includegraphics[angle=0,width=1.0\columnwidth]{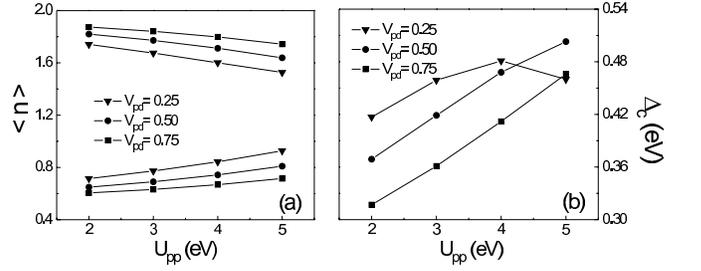}
\caption{(a) Occupation of the V\,$d_{xy}$ and O$_{\mathrm{R}}$\,$p_y$ orbitals. 
Up to the 2nd digit these numbers are constant along the ladder. 
For O$_{\mathrm{L}}$\,$p_x$ and O$_{\mathrm{L}}'$\,$p_y$ the occupation numbers are not less than 1.85. Results with $L\!=\!12$ rungs and $m\!=\!1600$ DMRG states. (b) Charge gap $\Delta_c$ for ladders with $L\!=\!12$ rungs and $3L$ holes, where 
$\Delta_c\!=\!1/2\,[E(3L\!+\!2)+E(3L\!-\!2)-2E(3L)]$\,.
The values extrapolated to $1/L\!\rightarrow\!0$ through $1/L\!=\!1/16$ 
and $1/L\!=\!1/20$ change by 15--25$\%$, with larger differences for higher $U_{pp}$\,.}
\end{figure}

Physical intuition would point to a strong electron\,--\,lattice
coupling for bond-stretching vibrations within the
V--\,O$_{\mathrm{R}}$--V unit. The situation should be somewhat
similar to the in-plane O half-breathing mode in doped copper
oxides, see for example \cite{el_ph_CuO_04} and Refs. therein. 
To study such
couplings in the MV ladder, we simulate the
displacement of the rung ligands by introducing a modulation of
the rung transfer integrals, $t_{r}\!\sim\!r^{-7/2}$
\cite{book_harrison}, and inter-site Coulomb repulsion,
$V_{pd}^{\,r}\!\sim\!1/r$. The effect of these displacements on
the GS (electronic) energy is investigated by DMRG
calculations.
However, before discussing the DMRG results, we illustrate the
nature of the electron\,--\,lattice interactions on an
oversimplified, single-rung model. We first
assume that in NaV$_2$O$_5$ three electrons (holes) are
``localized'' on each $d_{xy}$--\,$p_{y}$--\,$d_{xy}$ ``rung''.
For an isolated 3-site rung, we deduce then the expression for
the lowest energy eigenvalue corresponding to a superposition of
$p_{y\,}^{\,2}$ and $p_{y\,}^{\,1}$ configurations. With
$V_{pd}\!=\!0$, $\Delta_{\mathrm{eff}}\!=\!\Delta-U_{pp}$, and
neglecting double occupation at the $d$ sites, the Hamiltonian
reads in matrix notation
\begin{displaymath}
H_r = \left( \begin{array}{ccccc}
0        &0        &0                     &t_{R}                 &t_{R}\\
0        &0        &t_{L}                 &0                     &t_{L}\\
0        &\ t_{L}\ &\Delta_{\mathrm{eff}} &0                     &0  \\
\ t_{R}\ &0        &0                     &\Delta_{\mathrm{eff}} &0  \\
\ t_{R}\ &\ t_{L}\ &0                     &0 
                   &\Delta_{\mathrm{eff}}        \\
\end{array} \right),
\end{displaymath}
where $t_{L,R}(x)\!\approx\!t_{r}/(1\!\pm\!x)^{7/2}$,
$x\!=\!\delta x/r(\mathrm{VO}_{\mathrm{R}})$, $\delta x$ is the
displacement of the O$_{\mathrm{R}}$ ion along the rung direction, and
$r(\mathrm{VO}_{\mathrm{R}})\!\approx\!1.83$\,\AA \cite{navo_xrd_Pmmn}.
The lowest eigenstate is a doublet and has an energy
$E= 1/2\Delta_{\mathrm{eff}}-1/2\,(\Delta_{\mathrm{eff}}^2
   +4t_L^2+4t_R^2+4\,(t_L^4+t_R^4-t_L^2t_R^2)^{1/2})^{1/2}$.
We express $E(x)$ as an one-dimensional (1D) series expansion and
obtain:
\begin{eqnarray*}
E(x) = E(0) - \frac{1}{4}\,
              \frac{630\,t_{r}^{2}}
                   {\left(\,\Delta_{\mathrm{eff}}^2
                     + 12\,t_{r}^{2}\,\right)^{1/2} }\,x^2 + O(x^4) \ .
\end{eqnarray*}
We find thus a negative, ``electronic'' contribution to the harmonic
force constant associated with this displacement. If we take the
values from Table\,I, $t_{r}(0)\!=\!t_{r}\!=\!1.3$ and
$\Delta\!=\!3.5$, choose $U_{pp}\!=\!3$\,eV, express the
energy in units of meV and the distortion in
units of percents of the VO$_{\mathrm{R}}$ bond length, we find
$E(x)\!-\!E(0)\!\approx\!1/2\,k_{\mathrm{el}}x^2$,
$k_{\mathrm{el}}\!=\!-11.7$\,. This is a huge amount, about
35\,eV/\AA$^2$. For the isostructural CaV$_2$O$_5$ compound, a
$3d^1$ Mott insulator \cite{tanusri_cavo_00}, the infrared active
VO$_{\mathrm{R}}$ bond-stretching phonon was associated to the
$B_{3u}$ mode at 515\,cm$^{-1}$ \cite{phononsVO5_popo_02}, which
gives a force constant
$k'_{\mathrm{ir}}\!=\!15.6$\,eV/\AA$^2$. For NaV$_2$O$_5$\,, the
appearance of continuum features in this frequency range, in both
reflectivity $B_{3u}$ and Raman $B_{2g}$ spectra, makes the
interpretation difficult
\cite{phononsVO5_popo_02,phononsNaVO_popo99_basca00}.
Nevertheless, this continuum proves the existence of
strong electron\,--\,lattice interactions.

The large magnitude of this effect is confirmed by DMRG
calculations on the V--O Hubbard ladder, although more
complex interactions are now involved. 
Contributions to the harmonic force constant induced by 
electron\,--\,lattice couplings are plotted for various $U_{pp}$ and 
$V_{pd}$ values in Fig.\,2(a). The numbers were obtained by
introducing an in-line pattern of (static) V--\,O$_{\mathrm{R}}$--V 
distortions. Each O$_{\mathrm{R}}$ was displaced by the same amount and 
all other ions were kept fixed. 
To extract the $k_{\mathrm{el}}$ parameters we
calculated $E(x)$ for several small $x$ values.
We also considered the bending of the V--\,O$_{\mathrm{R}}$--V bond.

Effects induced by the vibronic mixing of two or several electronic
states are referred to as (pseudo) JT effects \cite{jt_gehring_bersuker}. 
The presence of quasilocalized and relatively close in energy
states plus the strong coupling between the electronic and nuclear
motion in NaV$_2$O$_5$ can be included then within this class of
phenomena. We checked for the existence of similar effects in
CaV$_2$O$_5$\,. CaV$_2$O$_5$ is also highly anisotropic
\cite{nmto_tpds_cavo}. However, the nature of the ground and
low-lying excited states is different and also their energy
separation is larger, not less than $U_{dd}$\,. Such interactions
are expected here to be weaker, which is indeed confirmed by DMRG
calculations on a ``half-filled'' $d^{1}$--\,$p^{6}$--\,$d^{1}$
ladder. For $U_{dd}\!=\!3$\,eV \cite{tanusri_cavo_00},
$2\!\leq\!U_{pp}\!\leq\!5$, and $0.25\!\leq\!V_{pd}\!\leq\!0.75$ the
electronic energy per single rung varies as
$E(x)\!-\!E(0)\!\approx\!1/2\,k'_{\mathrm{el}}\,\delta x^2$,
$2.4\!<\!-k'_{\mathrm{el}}\!<\!7.0$ (eV/\AA$^2$). Although less
drastic, this still should produce some phonon softening. We
obtain thus a ``bare'' harmonic force constant $k'= k'_{\mathrm{ir}} +
\mid\!k'_{\mathrm{el}}\!\mid$\,. 
We assume now that $k'$ is also a 
reasonable estimate for the bare harmonic force constant in 
NaV$_2$O$_5$ and display in Fig.\,2(b) an overall force
constant $A= k' - \mid\!k_{\mathrm{el}}\!\mid$ that should define the 2nd order term in the 1D Taylor expansion of the potential energy.
We see that for certain values of $U_{pp}$ and $V_{pd}$ $A$ is negative,
which shows that the system is instable with respect to such
deformations. The finite amplitude of these displacements is
related to the presence of higher order, positive terms. 
In the simplest approximation 
we arrive at the so-called $\phi^4$ model, widely used to model
local anharmonic effects in insulators \cite{book_bruce_cowley}:
$V_{R}\!=\! 1/2\,A\,x^2 + 1/4\,B\,x^4$.

%% FIGURE 2
\begin{figure}[t]
\includegraphics[angle=0,width=1.0\columnwidth]{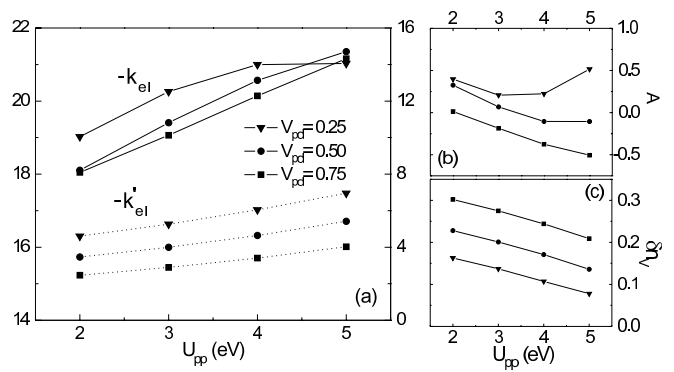}
\caption{(a) $k_{el}$ (full lines, left scale) and $k'_{el}$ (dots, right scale) parameters describing the effect of rung distortions on the electronic energy per rung of the V--O ladder in NaV$_2$O$_5$ and CaV$_2$O$_5$ (eV/\AA$^2$).
(b) $A$ parameter associated with the $\phi^4$ model, see text. Units of meV for energy and percents of the 
VO$_{\mathrm{R}}$ bond length for the amplitude of the distortion were used. 
(c) V--V charge disproportion
$\delta n_V$, $\delta n_V\!=\!n(d_{L,xy})\!-\!n(d_{R,xy})$,
for a zigzag arrangement of the rung oxygens
with  $x\!=\!\delta x/r(\mathrm{VO}_{\mathrm{R}})\!=\!\pm3\%$. 
In each figure $L\!=\!12$, $m\!=\!1600$, and the same symbols were used
to plot the curves.} 
\end{figure}

Near degeneracy and a large contribution of oxygen-hole
configurations to the GS wave-function were actually
predicted by multiconfiguration, quantum chemical calculations on
small clusters \cite{suaud_nav2o5_02,hozoi_nav2o5_03}. 
Even more, these
\textit{ab initio} calculations \cite{hozoi_nav2o5_03} indicate a
double-well potential for distortions of the rung oxygen along the
$x$ axis and give strong support for the model we propose here.
Whereas the approximate treatment of so-called dynamic electron correlation effects might somewhat
overestimate the weight of the O$_{\mathrm{R}}$\,$p_{y}^{1}$
configuration and cannot guarantee a highly accurate value for the 
$A$ parameter, we still can extract from the results of
\cite{hozoi_nav2o5_03} a good estimate for $B$,
essentially determined by ``hard'' core\,--\,core repulsion. 
Using 
the units of Fig.\,2(b) and
$V_{R}^{min.}\!=\!V_{R}(x_{0}\!\!\approx\!3\%)\!=\!-42$\,K
\cite{hozoi_nav2o5_03}, we find
$B\!=\!4\,V_{R}^{min.}/x_{0}^4\!=\!0.18$\,. 
The DMRG results show a 4th order term in the purely electronic $E(x)$ expansion of the order of $1\%$ of this value.
For the bond-stretching vibration of the O$_{\mathrm{R}}$ ion we can also 
evaluate inter-rung couplings. We model these oxygens by simple
point charges and consider only O$_{\mathrm{R}}$--\,O$_{\mathrm{R}}$
Coulomb interactions. To lowest order we find a n.n. interaction
\begin{eqnarray*}
V_{RR} &\approx &- \sum_{j=1}^{4}\frac{C_a}{2}\,(x_i - x_j^a)^2 \\
       &-       &\sum_{k=1}^{2}\frac{C_b}{2}\,(x_i - x_k^b)^2
               - \sum_{l=1}^{2}\frac{C_c}{2}\,(x_i - x_l^c)^2 \,,
\end{eqnarray*}
where $C_b\!=\!q^2 r(\mathrm{VO}_{{\mathrm R}})^2/{\it b}^3$ and
$C_c\!=\!q^2 r(\mathrm{VO}_{{\mathrm R}})^2/{\it c}^3$ are coupling
constants within the $bc$ plane and a slightly more complicated
expression corresponds to the inter-ladder coupling $C_a$ in the
$ab$ plane. With an ``effective'' charge for the O$_{\mathrm{R}}$ ion
$q\!\approx\!1.5$ and the lattice constants from
\cite{navo_xrd_Pmmn}, these inter-rung couplings are
$C_a\!=\!0.19$, $C_b\!=\!0.23$, $C_c\!=\!0.10$\,.
We note here
that the DMRG GS energy for a zigzag configuration of
the O$_{\mathrm{R}}$ oxygens is always slightly lower than the energy
for in-line displacements. This would give some extra contribution 
to the $C_b$ parameter.
Parametrized Hamiltonians based on low-order Taylor expansions of
the energy surface were intensively studied in connection with
structural phase transitions in ferroelectrics, by both
molecular-dynamics and Monte Carlo simulations
\cite{schneider_stoll_rev_80,kerr_79_86,batio_vanderbilt_95}.
Whereas such a numerical study is beyond the scope of this paper,
we still can make a rough estimate of the transition temperature.
In a mean field treatment $T_c\!\sim\!(2d\tilde{C})A/B$
\cite{book_bruce_cowley}, where $d$ is the system's dimension and
$\tilde{C}$ is an isotropic n.n. coupling. Although
not accurate, this gives at least the order of magnitude of
$T_c$\,, as comparison to numerical results shows
\cite{schneider_stoll_rev_80}. For $U_{pp}\!=\!5$ and
$V_{pd}\!=\!0.75$, $A\!=-0.51$ and with
$\tilde{C}\!\!\approx\!0.18$ we obtain $T_c\approx 47$\,K. We
note, however, that a number of other effects were neglected, such
as charge polarization, other $pd$ and $pp$ hoppings, longer range
electrostatic interactions, coupling to other modes,
\textit{e.\,g.} the V--\,O$_{\mathrm{R}}$--V bending, and coupling to
the strain. The last effects should
determine the actual LT crystal structure.

We turn now to the analysis of the GS wave-function with
displaced O$_{\mathrm{R}}$ ions and plot in Fig.\,2(c) the changes induced by
rung distortions on the V\,$d_{xy}$ orbital occupation numbers. 
For the $U_{pp}$ and $V_{pd}$ values
considered here, the on-rung V--V charge disproportion 
does not exceed 0.31 for 
$x\!=\!3\%$. The values would probably decrease somewhat when including on-rung polarization effects. 
Recent interpretations of the experimental data 
\cite{navo_delta_LT}
seem to agree with our results.

In conclusion, we investigated the interplay between electron and
lattice degrees of freedom for a MV ladder system. We
found that structural anisotropy and complex Coulomb correlations
lead to quasilocalized, three-hole states on each rung.
Charge fluctuations within such an ``unsaturated''
metal\,$3d$\,--\,O\,$2p$\,--\,metal\,$3d$ structure are coupled to
the lattice, which gives rise to strong JT
type interactions. We suggest that the phase transition in NaV$_2$O$_5$
is driven by cooperativity among these on-rung JT effects.
We mapped results of DMRG calculations for a
$p$\,--\,$d$ \,Hubbard type ladder onto a $\phi^{4}$ model. For
reasonable values of the on-site and inter-site electron repulsion
parameters we could reproduce in a mean field approximation the
order of magnitude of the transition temperature. The spin-gap
formation below 34\,K is probably related to the existence of
different V--\,O$_{\mathrm{L}}$--V paths between consecutive rungs on
the same ladder 
\cite{mostovoy_navo_00_02,bernert_spingap_01,hozoi_nav2o5_03}. 
The dielectric anomalies at $T_c$
\cite{navo_afe_99} should be induced by zigzag ion ordering and
only small V--V charge disproportion.
We also mention that for certain
values of the Coulomb repulsion parameters, see Fig.\,1, the 
GS wave-function has significant oxygen-hole character.
Strong contributions of oxygen-hole configurations were recently proposed for other MV oxides, such as the half-doped
manganites, see \cite{mv_co_coey_04} and Refs. therein. 

We thank J. Kortus, T. Saha-Dasgupta, M. Mostovoy, O. Jepsen, 
O. K. Andersen, and O. Gunnarsson for help at various
stages of this work and stimulating discussions. L.\,H.
acknowledges financial support from the Alexander von Humboldt
Foundation.


\begin{references}
\bibitem{2nd_note} The rung oxygen O$_{\mathrm{R}}$ has only two V nearest neighbors, whereas the leg oxygen O$_{\mathrm{L}}$ has three.
\end{references}
\end{document}